\documentclass[referee]{raa}        
\usepackage{graphicx}
\usepackage{times}
\usepackage{CJKutf8}
\usepackage[angle]{natbib}
\usepackage{epstopdf} 
\usepackage{amssymb} 
\usepackage{amsmath} 
\usepackage{verbatim}

\bibpunct{(}{)}{;}{a}{}{,}

\hypersetup{pdftitle = The title of my PDF, pdfauthor = My name, pdfsubject= The subject, pdfkeywords = keyword1 keyword2 keyword3} 
\hypersetup{colorlinks = true, linkcolor = green, anchorcolor = red, citecolor = blue, filecolor = red, pagecolor = red, urlcolor = red}

\begin{document}

\title{Statistical properties of radio flux densities of solar flares $^*$
\footnotetext{\small $*$ Supported by the National Natural Science Foundation of China.}
}

\volnopage{{\bf 20XX} Vol.\ {\bf X} No. {\bf XX}, 000--000}
\setcounter{page}{1}

\author{ Wang Lu\inst{1,2,3}, Liu Si-ming\inst{1,2,3}, Ning Zong-jun\inst{1,2,3}}

\institute{Purple Mountain Observatory, Nanjing 210000, China; {\it wanglu@pmo.ac.cn}\\
\and
	Key Laboratory of Dark Matter and Space Astronomy, Nanjing 210000, China\\                       
\and
	University of Science and Technology of China, Heifei 100049, Chian\\ 
\vs \no
{\small Received 20XX Month Day; accepted 20XX Month Day}
}

\abstract{Short timescale flux variations are closely related to the energy release process of magnetic reconnection during solar flares. Radio light curves at 1, 2, 3.75, 9.4, and 17 GHz of 209 flares observed by the Nobeyama Radio Polarimeter from 2000 to 2010 are analyzed with a running smooth technique. We find that the impulsive component (with a variation timescale shorter than 1 second) of 1 GHz emission of most flares peaks at a few tens of solar flux unit and lasts for about 1 minute and the impulsive component of 2 GHz emission lasts a shorter period and peaks at a lower flux level, while at the three high frequency channels the occurrence frequency of flares increases with the decrease of the flux density up to the noise level of the corresponding background. The gradual components of these emissions, however, have similar duration and peak flux density distributions. We also derive the power spectrum on different timescales and a normalized wavelet analysis is used to confirm features on short timescales. At a time resolution of 0.1 second, more than $\sim$ 60$\%$ of these radio light curves show significant flux variation on 1 second or shorter time scales. This fraction increases with the decrease of frequency and reaches $\sim$ 100$\%$ at 1 GHz, implying that short timescale processes are universal in solar flares. We also study the correlation between the impulsive radio flux densities and soft X-ray fluxes obtained with the GOES satellites and find that more than 65$\%$ of the flares with an impulsive component have their impulsive radio emission reach a peak value ahead of the soft X-ray fluxes and this fraction increases with the radio frequency.   
}

\authorrunning{Wang Lu et al.}	
\titlerunning{Supported by the National Natural Science Foundation of China.}
\maketitle

%
\section{Introduction}
\label{sect:intro}

It is generally accepted that solar flares are triggered by magnetic reconnection at small kinetic scales. The magnetic energy can then drive plasma heating,  particle acceleration, and large-scale bulk motion and waves in the flaring processes. Therefore the magnetic energy release processes of solar flares cover a large scale range both in space and time. In general, long timescale phenomena are related to large-scale MHD processes, while short timescale ones may originate from kinetic processes on small spatial scales. Energetic particles play important roles in these energy conversion processes since a significant fraction of the released magnetic energy is thought to be going into such particles. These energetic particles, especially electrons, are efficient emitters. Observations of emission produced by them constitute a major part of solar flare studies. In order to probe these multi-scale processes of magnetic energy conversion using light curves, one needs to analyse the corresponding power spectrum. Several methods have been developed to analyse power spectrum of light curves, including the classical wavelet analysis, Lomb-Scargle periodogram technique(\citealt{1982ApJ...263..835S,1982ApJ...263..854B}), and auto-correlation function, that can give robust estimate of the statistical significance of each spectral feature. 

To probe the process of magnetic reconnection, short timescale signals are the most relevant ones.  Short timescale flux variations have been observed in radio, optical, EUV, soft X-ray, hard X-ray, and even gamma-ray emissions of solar flares (\citealt{2005A&A...440L..59F, 2005ApJ...621..482V, 2014ARep...58..573K, 2015ApJ...807...72L, 2016ApJ...822....7K}). Flux variations below 1 second are thought to be directly related to the acceleration and/or propagation of energetic electrons at the flare site, mainly trajectories of electron beams in the corona (\citealt{2017ApJ...848...77W, 2018NatCo...9..146K}).  Radio emission can be produced by energetic electrons via both coherent and incoherent emission processes and is a generic component of solar flare emission. Radio observations also have the highest temporal resolution and are ideal for investigations of reconnection triggered processes at short timescales. The aim of this paper is to use the Nobeyama Radio Polarimeter (NoRP) observations of solar flares to explore short timescale properties of the radio flux densities. As a statistical study, we mainly use a relatively simple running smooth technique to analyse flux variations at short timescales.

This paper is organized as follows: In Section \ref{sect:Method}, we present our sample of solar radio flares and a running smooth technique is introduced to derive the power spectra of radio light curves. The emission is then separated into an impulsive and a gradual component. Statistical properties of these components are explored.  In Section \ref{sect:Analysis}, the power spectra are used to study emission at short timescales and wavelet analyses are used to verify our results. In Section \ref{sect:discussion} we present the correlation between the radio fluxes and X-ray fluxes obtained with the GOES satellites. Our conclusions are drawn in Section \ref{sect:conclusion}.

\section{The Sample, Analysis Method, and Statistical Properties of the Impulsive and Gradual Components}
\label{sect:Method}

\subsection{The Sample}

The present study is based on a list of radio bursts recorded with the Nobeyama Radio Polarimeter (NoRP) from 2000 to 2010 at multiple frequencies: 1GHz, 2GHz, 3.75GHz, 9.4GHz, and 17GHz (\citealt{1985PASJ...37..163N}). The raw data of the NoRP have a time resolution of 0.1s. For each day of the recorded bursts, we selected two quiescent periods before and after each flare and calculated their mean flux densities and their standard deviations for each frequency. We redefine the start and end times of a flare at each frequency as the first and last five consecutive points whose flux densities exceed the mean quiescent flux densities by more than five times the corresponding standard deviation, respectively. To ensure that the signal is strong enough for spectral analysis, we only consider flares with a duration greater than 200 seconds and a peak flux density at least 1.1 times the corresponding averaged quiescent flux densities at one of these frequencies. Then 209 flares were selected. For our analyses, we also requires that the peak flux density at each frequency exceeds the corresponding averaged quiescent flux densities by more than 10\%, and some frequencies are dropped due to calibration of the flux density during the flare. Then at each frequency the number of bursts is less than 209. Overall, we have 182, 195, 200, 197, and 166 events for the 1GHz, 2GHz, 3.75GHz, 9.4GHz, and 17GHz frequency channels, respectively.

\subsection{Power Spectral Analysis}
\begin{figure}[htb]
\begin{minipage}{15cm}
\centering
\includegraphics[width=15.0cm, angle=0]{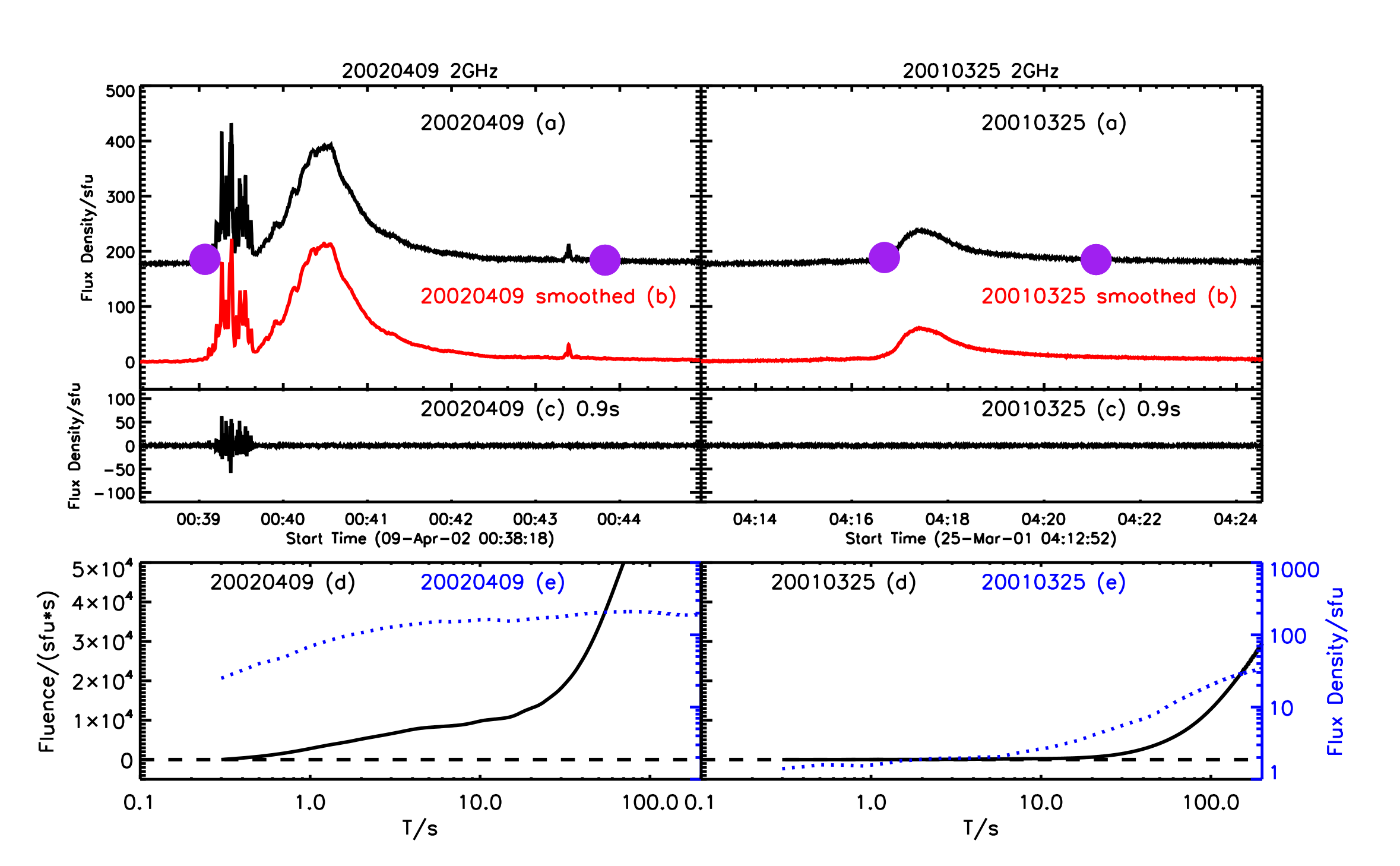}
\end{minipage}
\caption{Radio light curves of two flares (a), their 0.9 second smoothed background subtracted counter parts (b), the difference between (a) and their 0.9s smoothed counter parts (c), and their power spectra calculated with equation \ref{Pspec1} (d). (e) shows the maximum of the difference between the original and smoothed light curves at different timescales. The purple dots on (a) indicate the start and end times of these flares.} 
\label{Fig0}
\end{figure}
To quantify the significance of flux variation on different timescales, we use a running smooth technique. To subtract contribution from the background properly, we selected background periods at least 5 minutes away from the flare's start and end points at each frequency. The upper curves (a) in the top panels of Figure \ref{Fig0} show light curves of two flares at 2GHz. The lower curves (b) in these panels show the background subtracted light curves running smoothed over a duration of 0.9s. The curves (c) in the middle panels show the difference between (b) and (a) with the background subtraction. One can define the power of the flux density variation at 0.9s as the sum of the absolute values of (c). To subtract contributions to the power spectrum from the background, the background power spectra are also calculated and rescaled before the subtraction.
Then we may have the power at ${\rm T}=t \times 0.1$ s:
\begin{equation}
F({\rm T})={\sum_i|f_i-f_i{(t)}|}-{\sum_j|b_j-b_j{(t)}|\times {{l}\over{l_{b}}}}\,,
\label{Pspec0}
\end{equation}
where $f_i$ and $f_i(t)$ are the original and smoothed flare flux densities, respectively, $b_j$ and $b_j(t)$ are for the background, and $l$ and $l_b$ represent the duration of the flare and the background, respectively.

\begin{figure}[htb] 
\begin{minipage}{15cm} 
\centering
\includegraphics[width=15.0cm, angle=0]{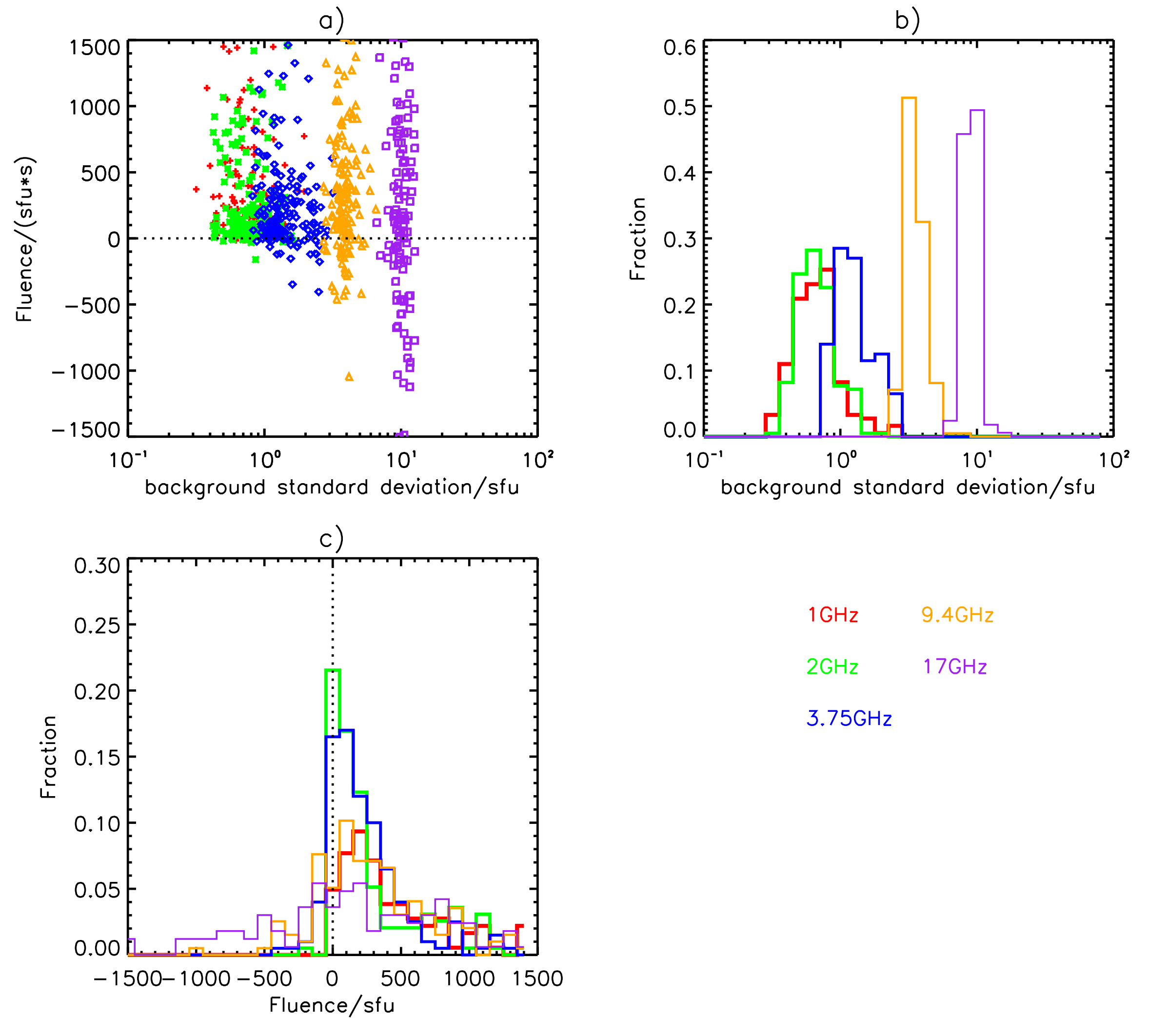}
\end{minipage}
\caption{a) Scatter plot of the power at 0.3 s and the standard deviation of the corresponding background period of each light curve at different frequencies. The corresponding normalized distributions of these flares are given in panels b (for the standard deviation) and c (for the power at 0.3s).} 
\label{Fig1}
\end{figure}

In general, the power spectra should always be positive. However, due to the background subtraction, negative power may be obtained at some timescales. Figure \ref{Fig1} shows the scatter plot of the power at 0.3 s indicated as fluence v.s. the standard deviation of the background (panel a) and their normalized distributions (panel b and c). With the increase of frequency, there are more fluctuations in the background, and more events have a negative power at 0.3s. 

To avoid uncertainties close to the temporal resolution of 0.1 s, one may set the power at 0.3 s to be zero with the background power spectrum normalized at T $= 0.3$ s. Then we may have the power at ${\rm T}=t \times 0.1$s: 
\begin{equation}
F({\rm T})={\sum_i|f_i-f_i(t)|}-\sum_i|f_i-f_i{(3)}|\times  \left[{\sum_j|b_j-b_j(t)|}/ \sum_j|b_j-b_j{(3)}|\right]\,.
\label{Pspec1}
\end{equation}
The solid lines in the bottom panels of Figure \ref{Fig0} show these power spectra. Comparing the two flares in Figure \ref{Fig0}, one can see that the flare on April 9, 2002 shown in the left panel has prominent impulsive emission, and its power spectrum starts to rise at the shortest time scale. While there is no evident impulsive emission for the flare on March 25, 2001 shown in the right panel and its power spectrum starts to rise after 10 seconds. 

In the bottom panels, we also show the maximum difference between the original and smoothed light curves $f_t$ at different timescales, which can be interpreted as the maximum contribution to the flux density from the impulsive component. The size of the emission region has an upper limit of the light travel distance $L_t\le t\times 0.1$s $\times c$, where $c$ is the speed of light. Then the brightness temperature on different timescales has a lower limit of:
\begin{equation}
T_b({\rm T})\ge f_t\left(D^2\over L_t^2\right)\left(c^2\over 2\nu^2k_{\rm B}\right) \simeq 0.82\left(0.1 t\nu\over 1\ {\rm GHz}\right)^{-2}\left(f_t\over 10\ {\rm sfu}\right) {\rm MK} 
\label{brightT}
\end{equation}
where $k_{\rm B}$ is the Boltzmann constant, $D=1.5\times 10^{13}$ cm is the distance to the Sun, $\nu$ is the observation frequency, and $1 {\rm sfu} = 1.0\times 10^{-19}$erg s$^{-1}$cm$^{-2}$Hz$^{-1}$. When the impulsive emission is prominent, this figure shows that the brightness temperature has a characteristic value higher than the coronal temperature and this temperature increases with the decrease of timescale T. The brightness temperature of the gradual emission is at least two orders of magnitude lower.

\subsection{Statistical Properties of the Impulsive and Gradual Components}

\begin{figure}[htb]
\begin{minipage}{15cm}
\includegraphics[width=15cm]{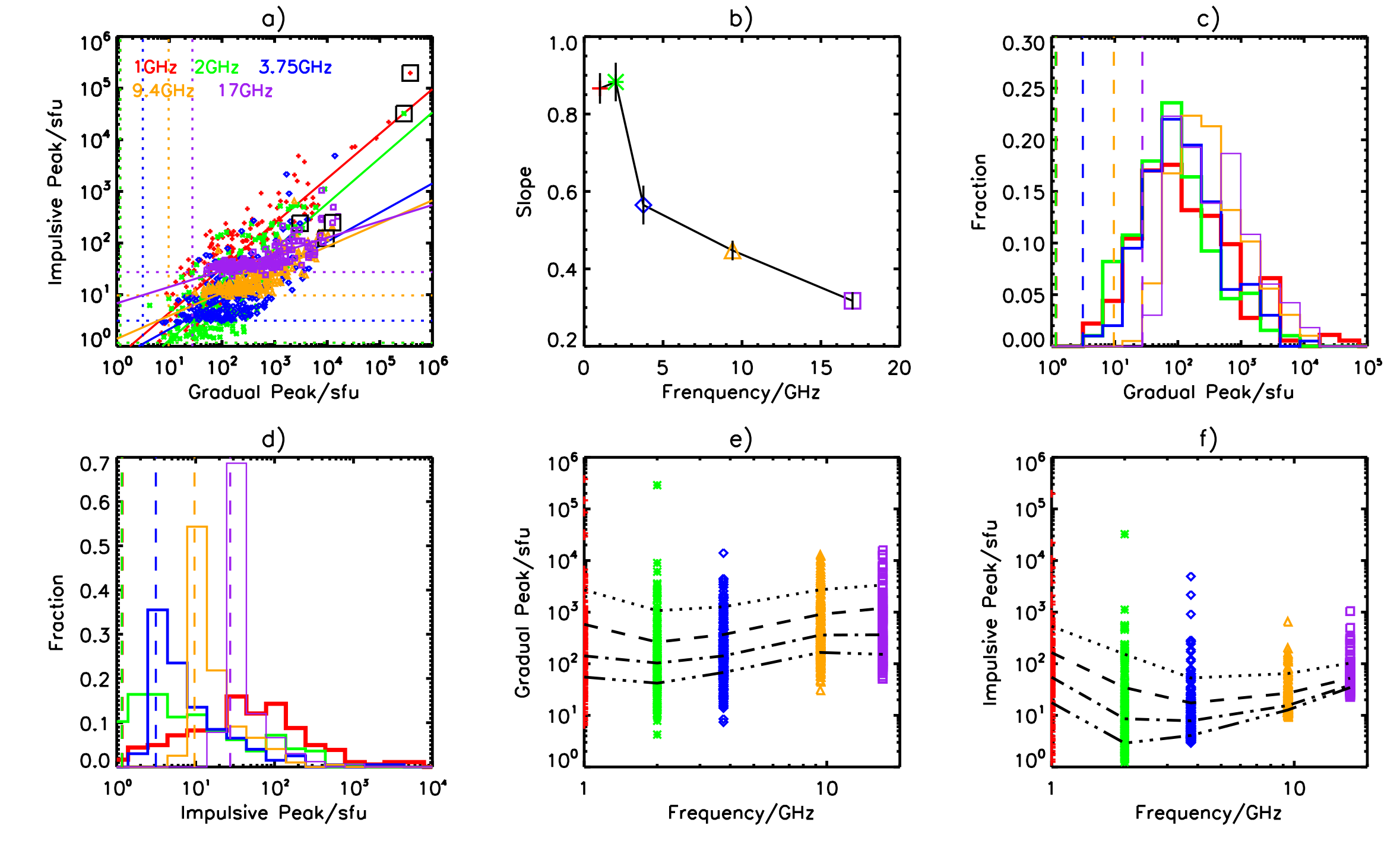}
\end{minipage}
\caption{a: Correlation between the peak flux densities of the impulsive and 0.9 second smoothed components. The dotted lines indicate flux densities that are 5 times the averaged standard deviation of the impulsive components of the corresponding backgrounds; b: Frequency dependence of the slope of the peak flux density correlation in Panel a; Occurrence frequency distribution of the peak flux densities of the smoothed (c) and impulsive (d) components. The dashed lines are the same as the dotted lines in Panel a; Scatter plots of the peak flux densities for the smoothed (e) and impulsive (f) components. From top to bottom, the lines indicate an event fraction of 10\% (dotted), 25\% (dashed), 50\% (dotted-dashed), and 75\% (triple-dotteddashed), respectively.}
\label{Fig2}
\end{figure}

Since phenomena below 1 second are closely related to the electron acceleration and transport processes, one may use T$=0.9$ s to separate the light curves into an impulsive (c of Figure \ref{Fig0}) and a gradual (b of Figure \ref{Fig0}) component.  Panel a of Figure \ref{Fig2} shows the correlation between the peak of the gradual component, i.e. the peak flux density of the background subtracted smoothed light curves, and the peak flux density of the impulsive component, the maximum difference (positively defined) between the original and the smoothed light curve. Panel b of Figure \ref{Fig2} shows that the slope of this correlation can be divided into two groups, one for the two low-frequency channels and the other for the three high-frequency channels. As shown in the left panel of Figure \ref{Fig0}, the peaks of the gradual components at the two low frequencies can have contributions from many simultaneous short timescale pulses and do not necessarily reflect physical processes on large scales. This may explain the good correlation between the peak flux densities of the impulsive and gradual components at 1 and 2 GHz. The slope decreases with the increase of frequency at high frequencies, where the impulsive component is less prominent. 

Panels c and d of Figure \ref{Fig2} show the normalized distributions of the peak flux densities of the gradual and impulsive components, respectively. At 1 GHz, the distributions of the gradual and the impulsive component are very similar, implying contamination of the gradual component by many impulsive pulses, and both peak at a few tens of sfu. At other frequencies, while the gradual components peak near 100 sfu, the impulsive components peak near the level of the background fluctuations, which is defined as 5 times of the averaged standard deviation of the impulsive component of the corresponding backgrounds, implying more flares with lower impulsive peak flux densities. 

Panels e and f of Figure \ref{Fig2} show the scatter plots of the peak flux density distributions of the gradual and impulsive components, respectively. Here again one can see that these distributions at 1 GHz are similar, but there are quantitative differences. For most flares, the peak flux density of the gradual component is higher than that of the impulsive component (See panel a), and the distribution of the gradual component appears to shift toward a slightly higher flux density. In particular, the medium flux density is above 100 sfu for the gradual component but less than 100 sfu for the impulsive component. 
These results suggest that the gradual component at 1 GHz is usually not as strong as the impulsive component, and the gradual component we derived above can have significant contributions from pileups of many impulsive pulses. If this pileup effect does not depend on the peak flux density, as implied by self-similarity of X-ray fluxes of solar flares(\citealt{2015AcASn..56...35Z}), one expects an identical peak flux density distribution of the impulsive and gradual components defined above. 

At the three high frequency channels, most of the impulsive components have a peak flux density comparable to the background fluctuations (Panels d and f), while the peak flux density distribution of the gradual components may be approximated by log-normal distributions (Panels c and e). At 2 GHz the distribution of the peak flux density of the gradual component is contaminated by the impulsive emission (Panel a and the left panel of Figure \ref{Fig0}), while the distribution of the peak flux density of the impulsive component is affected by the background (Panel d and f). Using equation \ref{brightT}, one can estimate the lower limit to the maximum brightness temperature from the impulsive peak flux densities at different frequencies. At 1 GHz, the brightness temperature can reach a few tens of billion Kelvin. This temperature decreases by more than four orders of magnitude above 10 GHz, implying distinct emission processes.

\begin{figure}[h]
\begin{minipage}{15cm}
\center{\includegraphics[width=15cm]{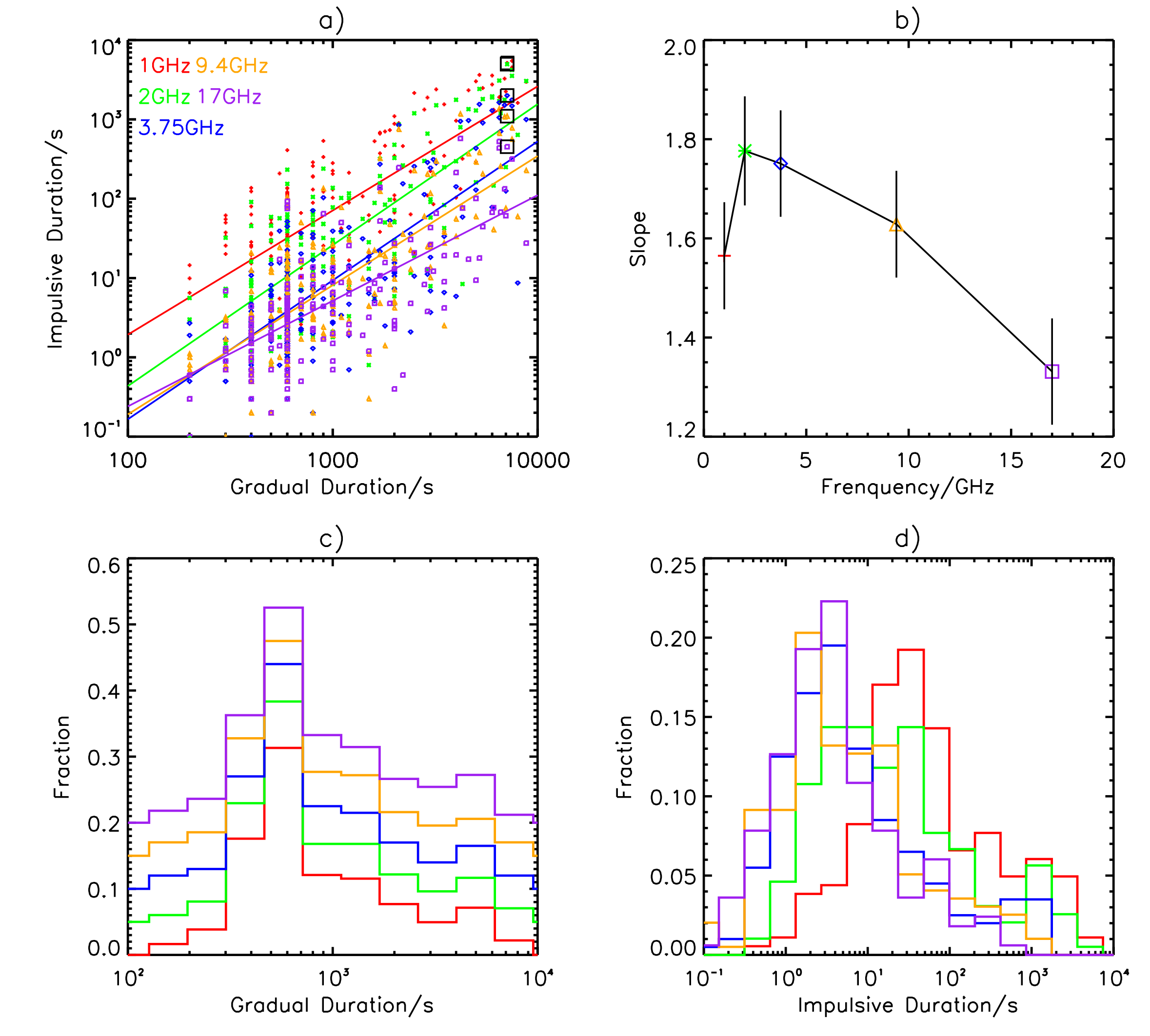}}
\end{minipage}
\caption{a: Correlation between the duration of the smoothed and impulsive components; b: Frequency dependence of the slope of the duration correlation in Panel a; Occurrence frequency distributions of the duration of the smoothed (c) and impulsive (d) components. For illustrative purposes, in Panel c, these distributions are shifted upward by 0.05 successively from top to bottom.}
\label{Fig3}
\end{figure}

We also extracted the intervals with a flux density above the background fluctuations defined as 5 times the standard derivation of the impulsive component of the corresponding backgrounds. The durations of the impulsive and gradual components are defined as the total duration of these corresponding intervals. Figure \ref{Fig3} shows statistical properties of these durations. The duration of the impulsive component is always lower than that of the corresponding gradual component and decreases with the increase of frequency (Panels a and d of Figure \ref{Fig3}). However the duration of the impulsive component increases quickly with the increase of that of the gradual component and these two become comparable at 1 and 2 GHz for flares with long impulsive durations (Panels a and b of Figure \ref{Fig3}). This again shows that the gradual components at low frequencies are contaminated by the impulsive emission. 

It is interesting to note that the distributions of durations of the gradual components are nearly identical at all frequencies (Panel c of Figure \ref{Fig3}), which implies a physical correlation between the impulsive and the gradual emission. As mentioned above, the gradual component at 1 GHz has significant contributions from pileups of many impulsive pulses. A physical correlation between the impulsive and the gradual emission can naturally explains that it has the same duration distribution as the other frequencies. We also note that radio emission is well-correlated with the impulsive phase of X-ray flares(\citealt{1968ApJ...153L..59N})and the gradual component studied here is mostly emitted in the impulsive phase of solar flares. However, the distribution of durations of the impulsive components at 1 GHz peaks at a few tens of seconds, which is one order of magnitude longer than peak locations of the distributions of durations at the three high frequencies (Panel d of Figure \ref{Fig3}). Again, the durations of the impulsive components at these high frequencies are likely affected by background fluctuations and do not reflect the intrinsic properties of the impulsive emission. The distribution of the durations of the impulsive component at 2 GHz is broader than that at other frequencies (Panel d of Figure \ref{Fig3}), reminiscent of the distribution of the corresponding peak flux density (Panel d of Figure \ref{Fig2}).

\begin{figure}[h]
\begin{minipage}{15cm}
\centering
\includegraphics[width=15cm]{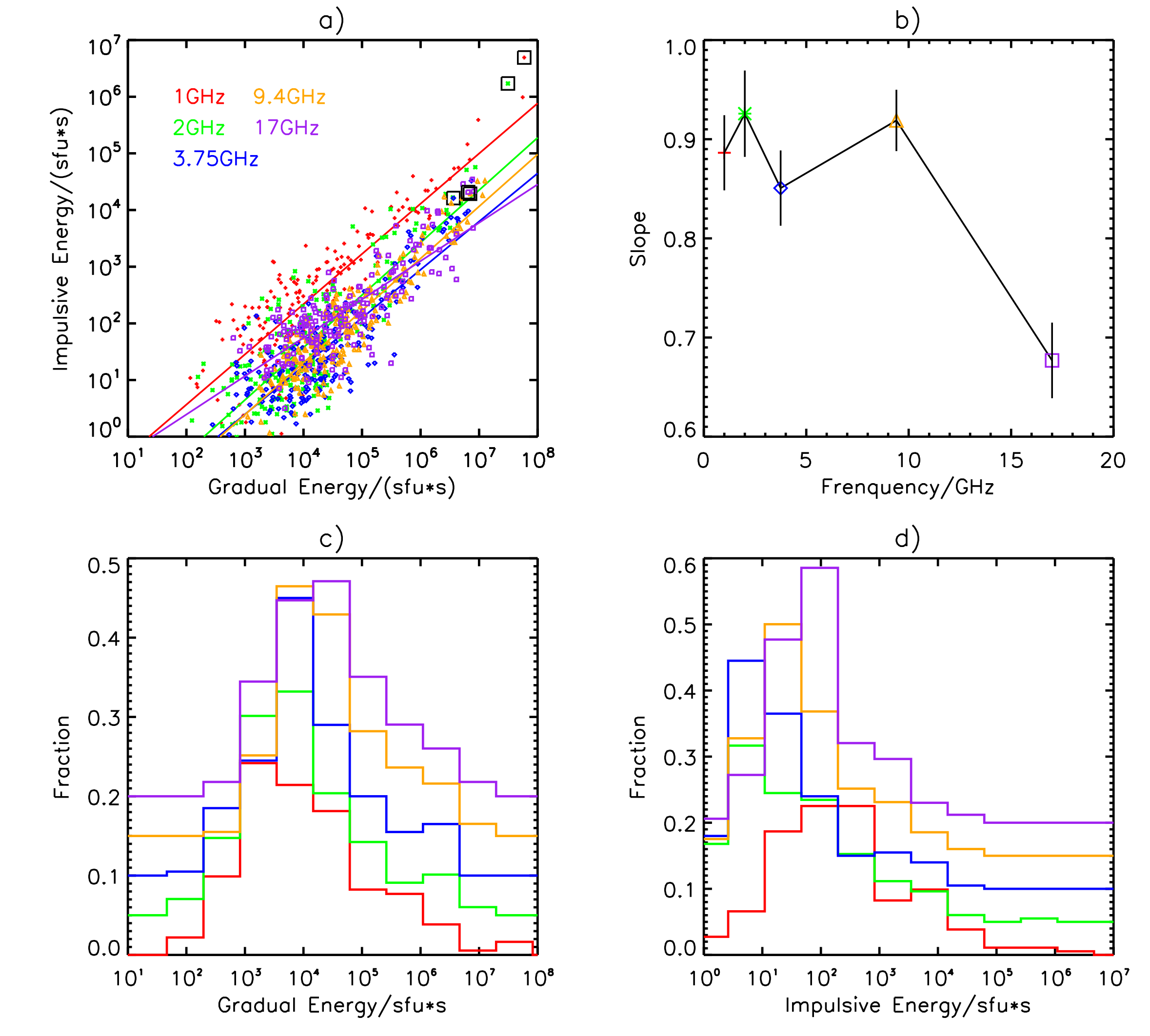}
\end{minipage}
\caption{Same as Fig. \ref{Fig3} but for the energy densities. From top to bottom, the distributions in Panels c and d are shifted upward by 0.05 successively.}
\label{Fig4}
\end{figure}

\begin{figure}[bht]
\begin{minipage}{7.0cm}
\centering
\includegraphics[width=15cm, angle=0]{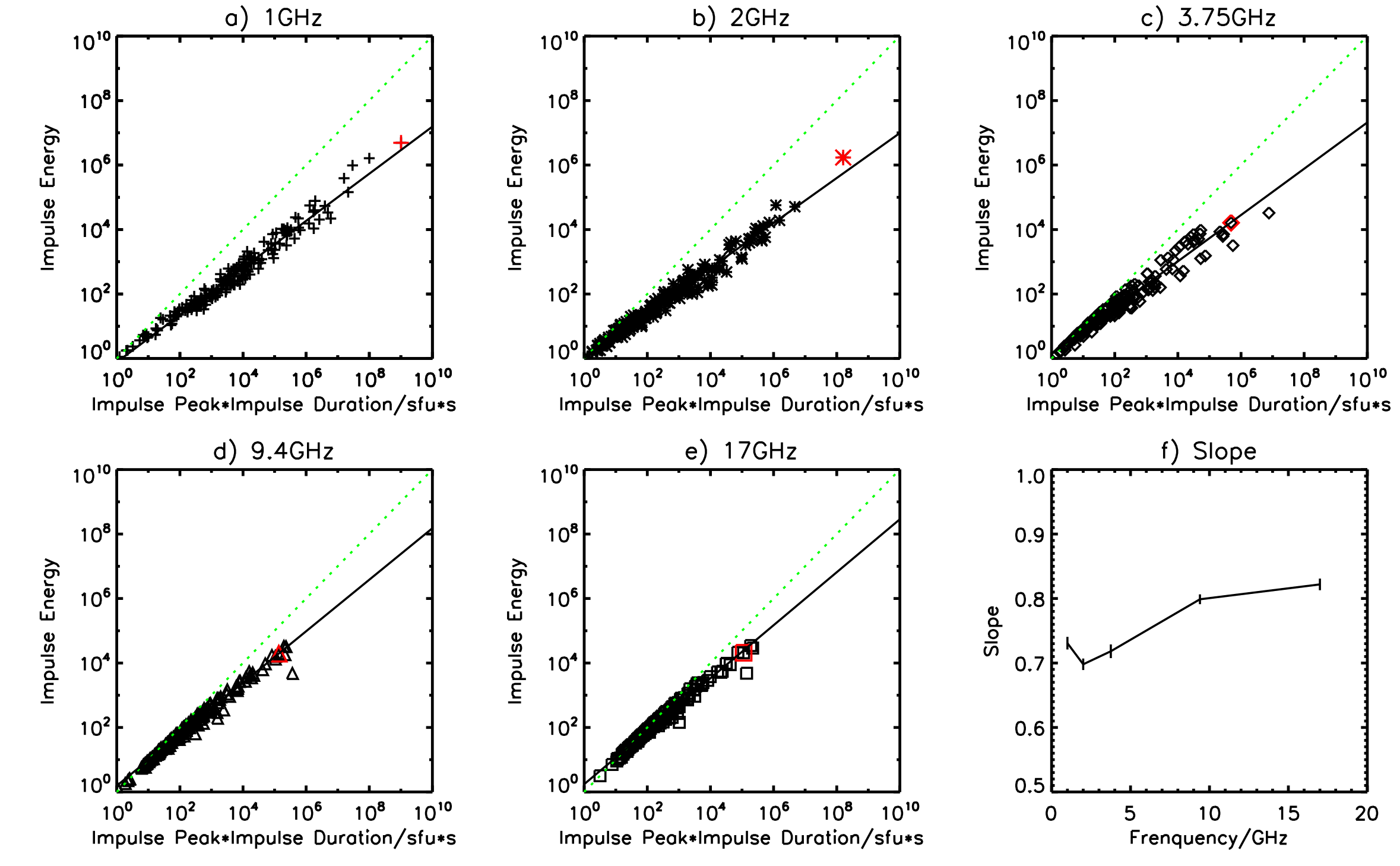}
\end{minipage}

\begin{minipage}{7.0cm}
\centering
\includegraphics[width=15cm, angle=0]{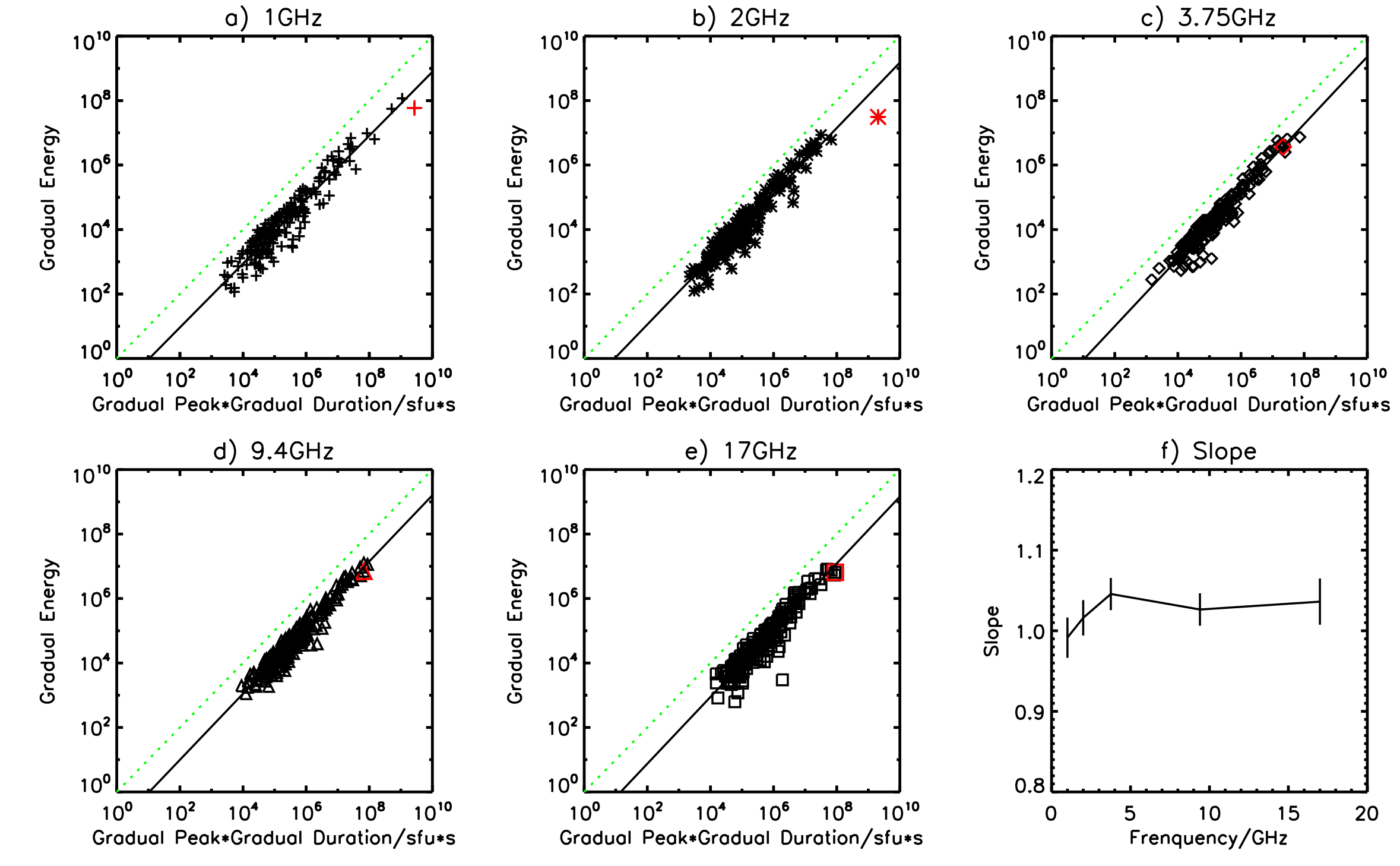}
\end{minipage}
\caption{Correlation between the energy densities and the product of  duration and peak flux densities. The top and bottom two rows are for the impulsive and gradual components, respectively. The f panels show the frequency dependence of the slopes of these correlations.} 
\label{Fig5}
\end{figure}

We also integrated flux densities of the impulsive and gradual components over the duration to obtain the corresponding energy densities (i.e., fluences). Figure \ref{Fig4} shows their statistical properties. The fluence of the impulsive component is always lower than that of the gradual component (Panel a) and the two components appear to be well correlated except for the 17 GHz channel (Panels a and b), where the background fluctuation is the strongest. Panel b shows that the slopes of these correlations are always less 1, implying relatively less impulsive emission in more powerful flares. Interestingly, peak locations of the fluence distributions of the gradual components appear to increase monotonically with the increase of frequency (Panel c), while those of the impulsive components decrease from 1GHz to 2 GHz first and then increase monotonically with the frequency (Panel d), implying distinct emission mechanisms at 1 GHz. In general, most of the energy, which can be estimated with the product of the energy density and the frequency, is radiated via the gradual component at high frequencies.

To further draw distinctions between the impulsive and gradual components, in Figure \ref{Fig5} we show the correlations between the fluence and the product of duration and peak flux density of the impulsive (top two rows) and gradual components (bottom two rows). In general, these quantities are well correlated. However, the slopes for the gradual component are always close to 1 (Panel f of the bottom two rows), as expected for the relatively smooth and seemingly energy independent light curves. The slopes for the impulsive components, however, are always less than 1 (Panel f of the top two rows), implying shorter durations or less frequent appearance of stronger pulses. These two quantities become comparable for small events and the slope appears to be higher at the two high frequencies, implying different origin of the impulsive emission at different frequencies. Nevertheless, the power-law scaling of these two quantities still implies an energy (or scale) independent behaviour of the impulsive component. In summary, these correlations imply that most energy of the gradual components is released near the peak of the flux density, while for the impulsive component, most energy is released by the more numerous pulses with a lower flux density.

\section{Variation of Flux Densities at Short Timescales}
\label{sect:Analysis}

In this section, we will have more quantitative study of flux density variation on short timescales.

\begin{figure}[htb] 
\begin{minipage}{15cm}
\centering
\includegraphics[width=15.0cm, angle=0]{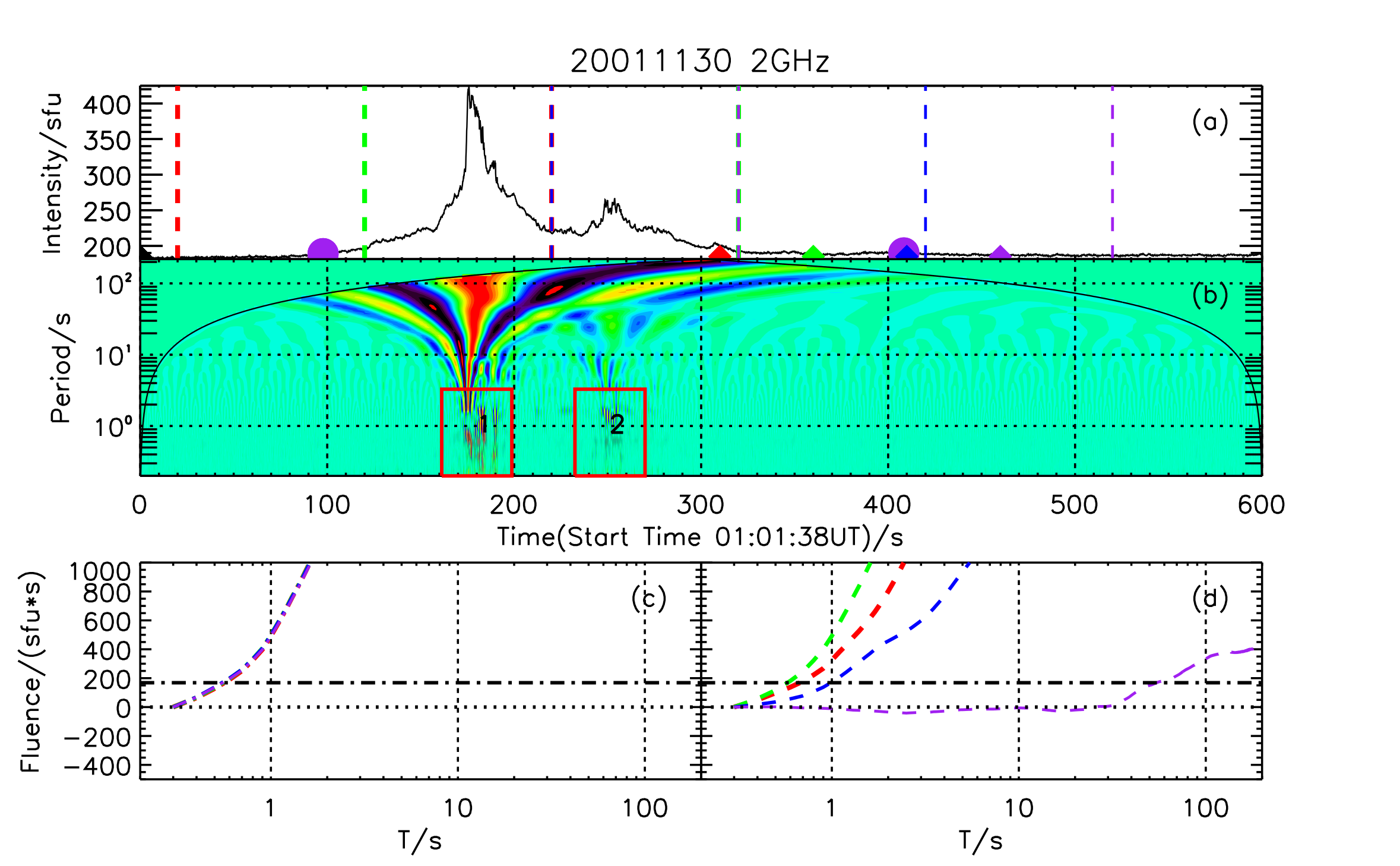}
\end{minipage}
\caption{Power spectra of selected periods of an event. Panel c is for periods starting at the black triangle and ending at triangles of different colors in Panel a. These lines essentially overlay with each other. Panel d is for equal length of periods indicated by dashed lines with different colors in Panel a. The black dotted-dashed lines in Panels c and d indicate a selection threshold for significant short timescale variations. See text for details.} 
\label{Fig6}
\end{figure}

\subsection{Normalized Wavelet Analysis}

\begin{figure}[htb] 
\begin{minipage}[]{7.5cm}
	\includegraphics[width=7.5cm, angle=0]{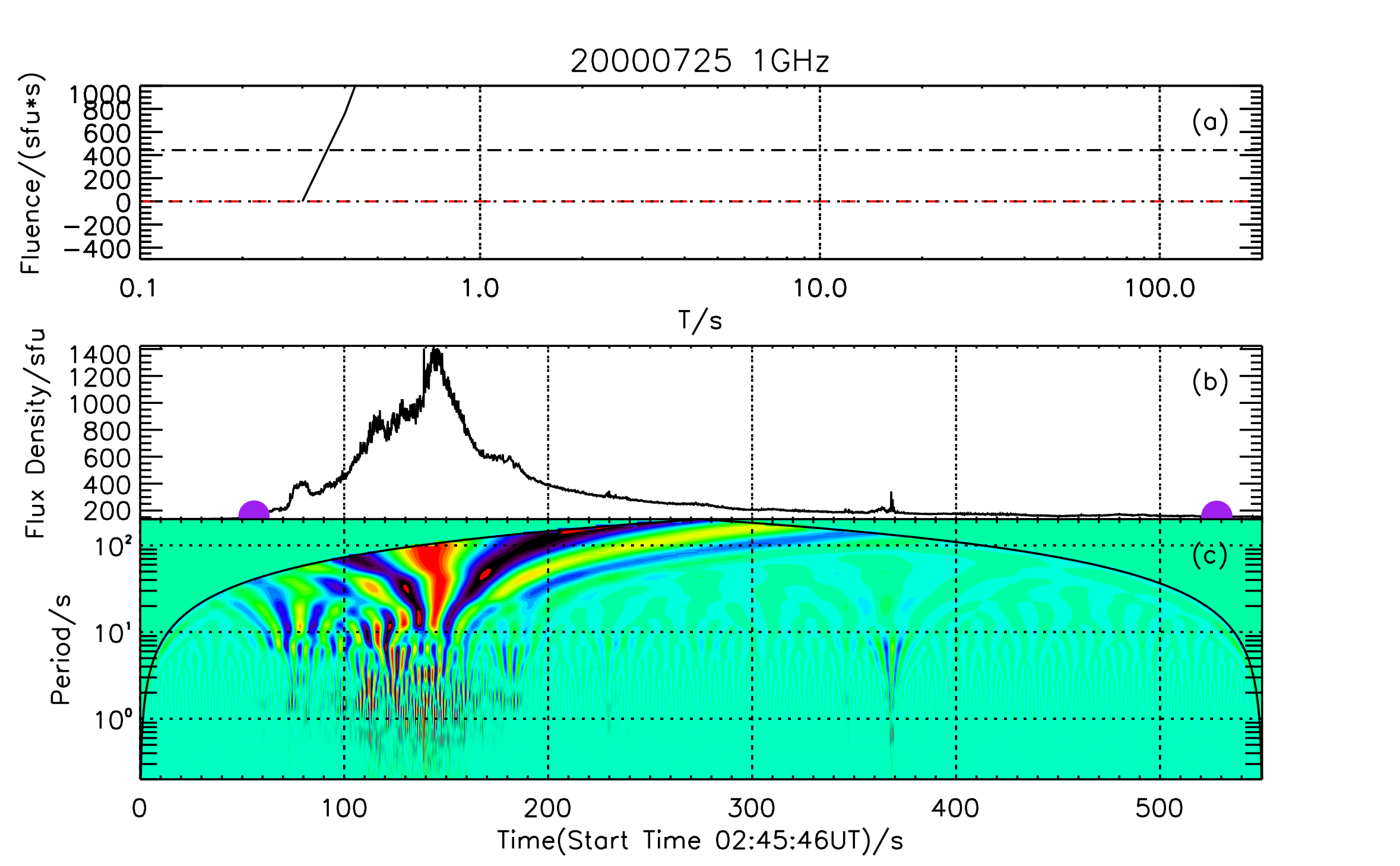}
\end{minipage}
\begin{minipage}{7.5cm}
	\includegraphics[width=7.5cm, angle=0]{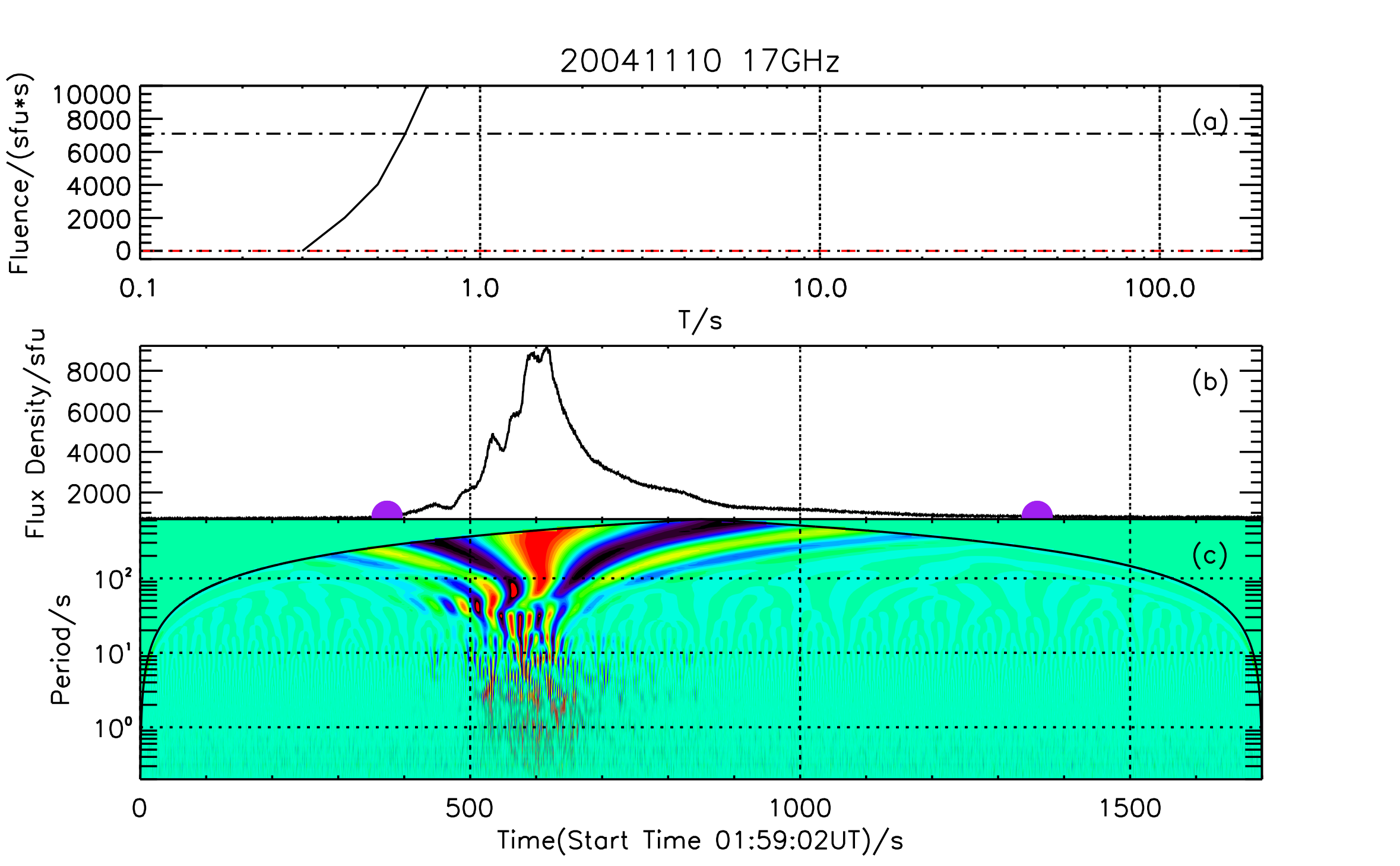}
\end{minipage}
\begin{minipage}{7.5cm}
	\includegraphics[width=7.5cm, angle=0]{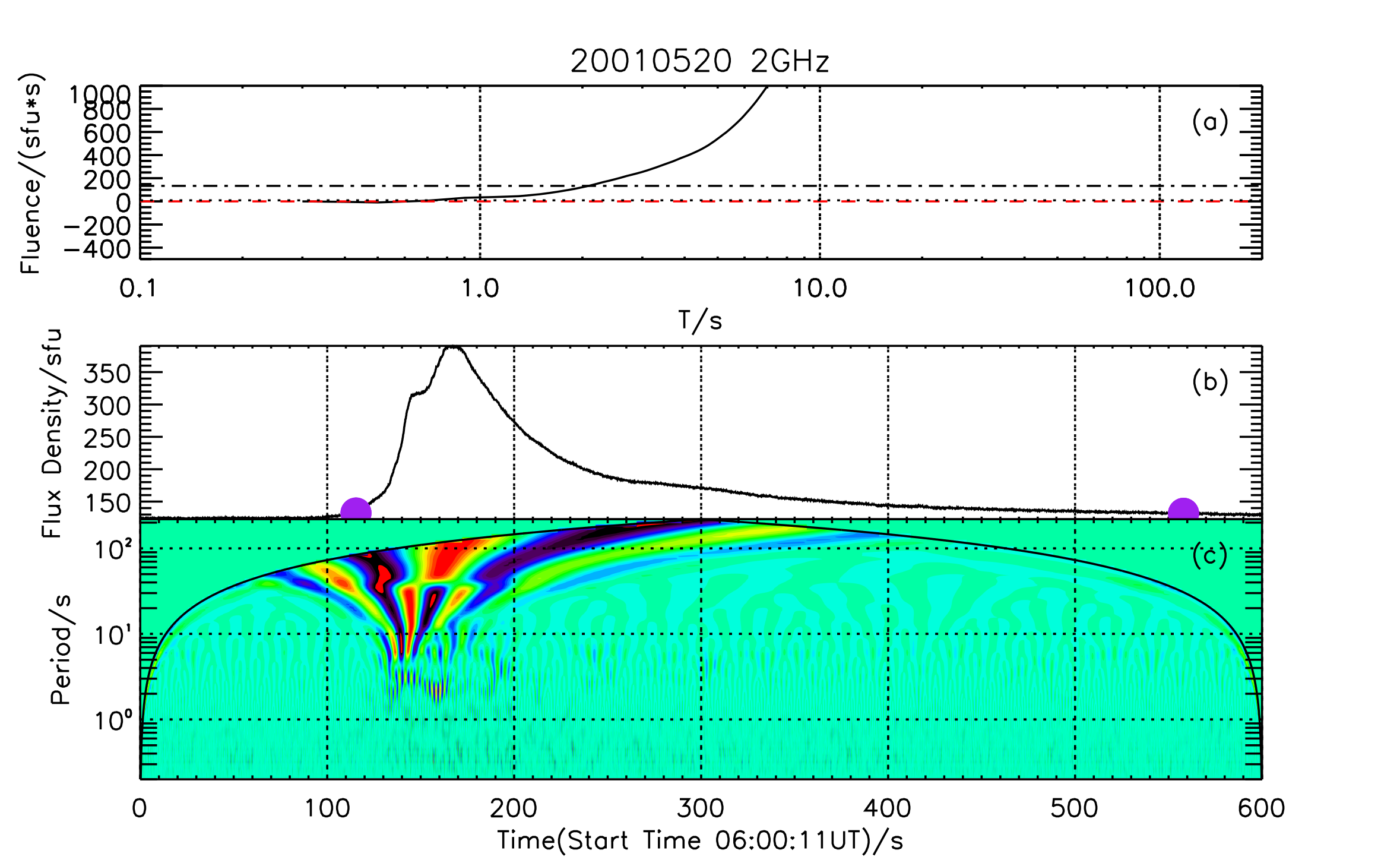}
\end{minipage}
\begin{minipage}{7.5cm}
	\includegraphics[width=7.5cm, angle=0]{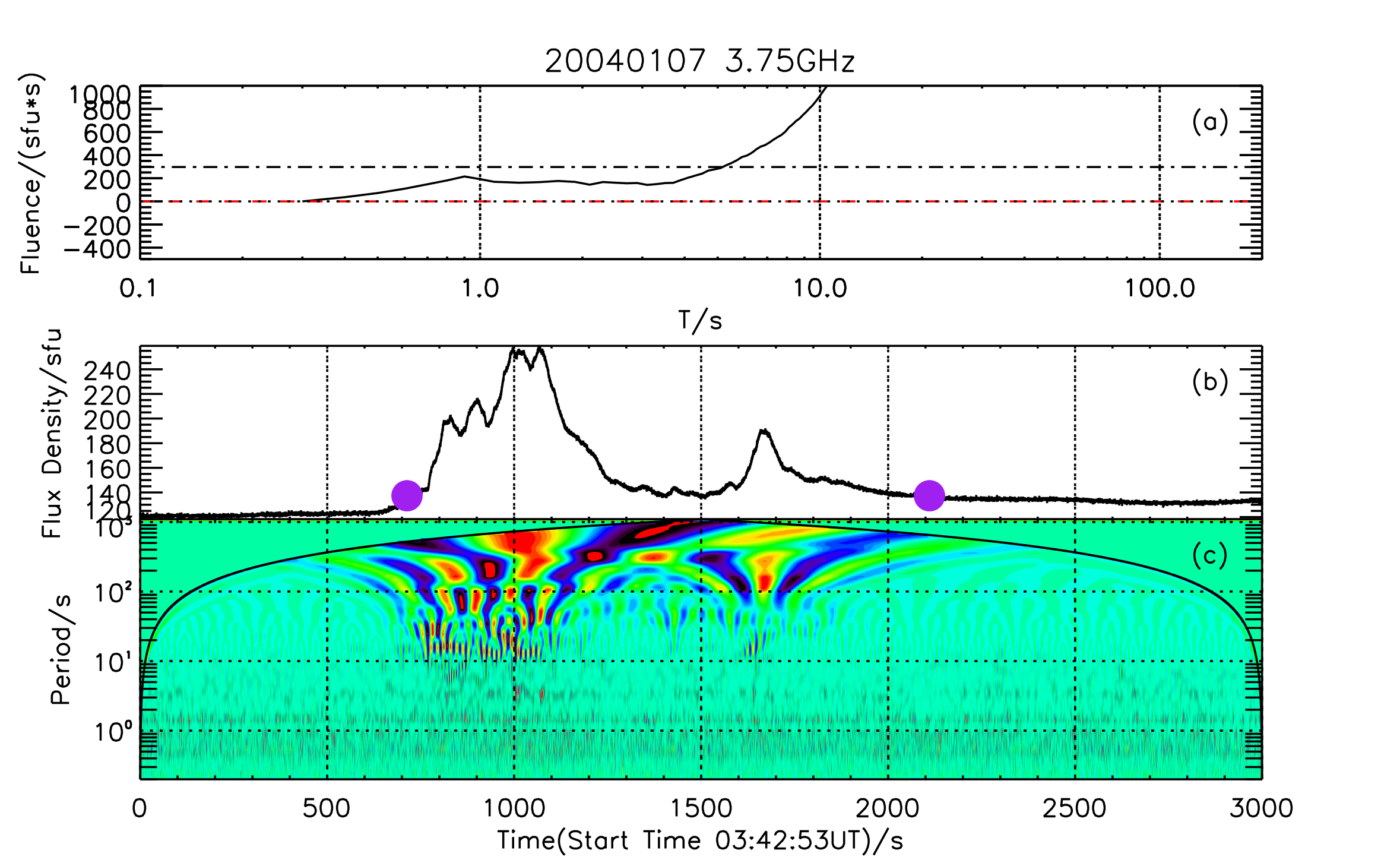}
\end{minipage}
\begin{minipage}{7.5cm}
	\includegraphics[width=7.5cm, angle=0]{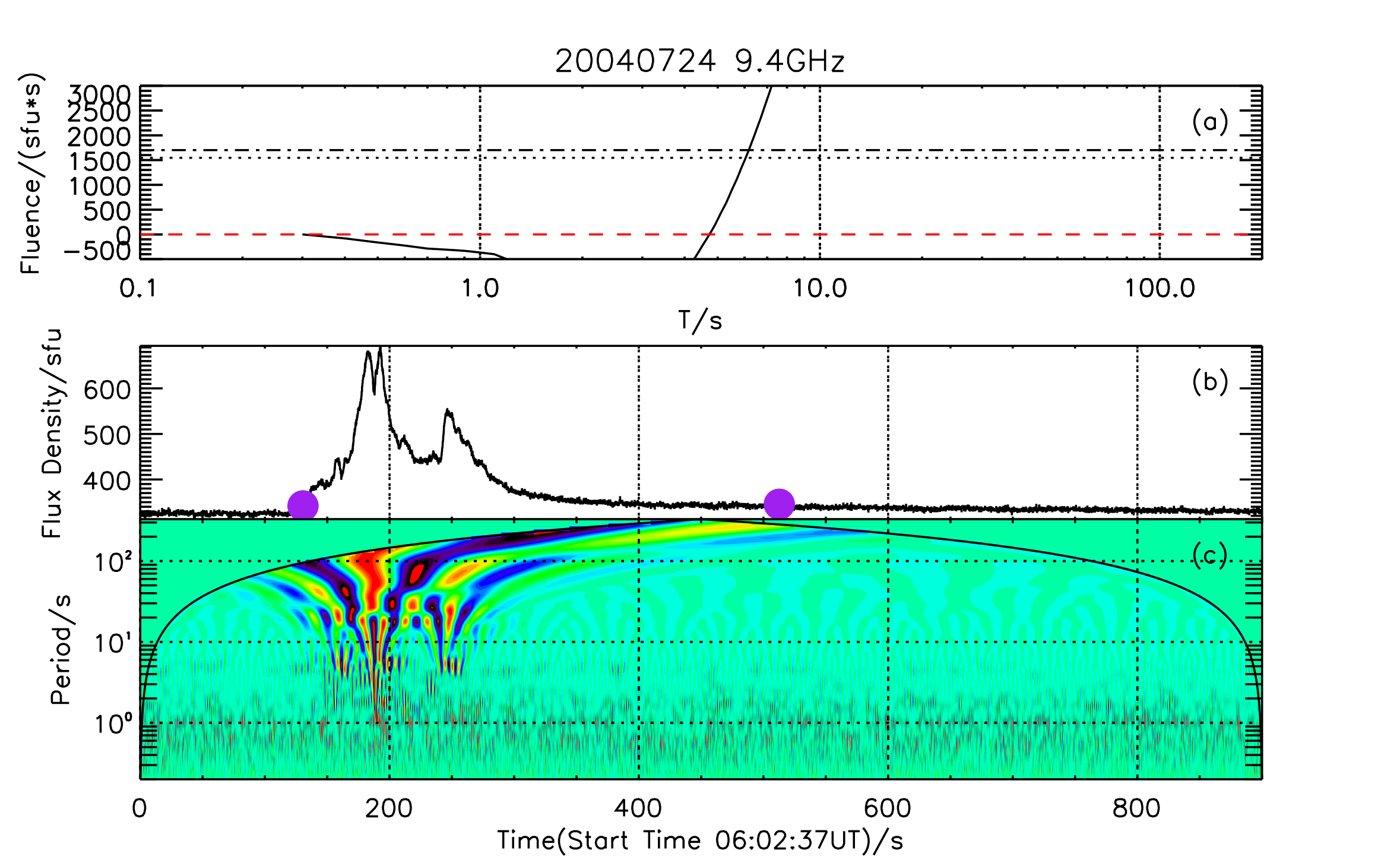}
\end{minipage}
\begin{minipage}{7.5cm}
	\includegraphics[width=7.5cm, angle=0]{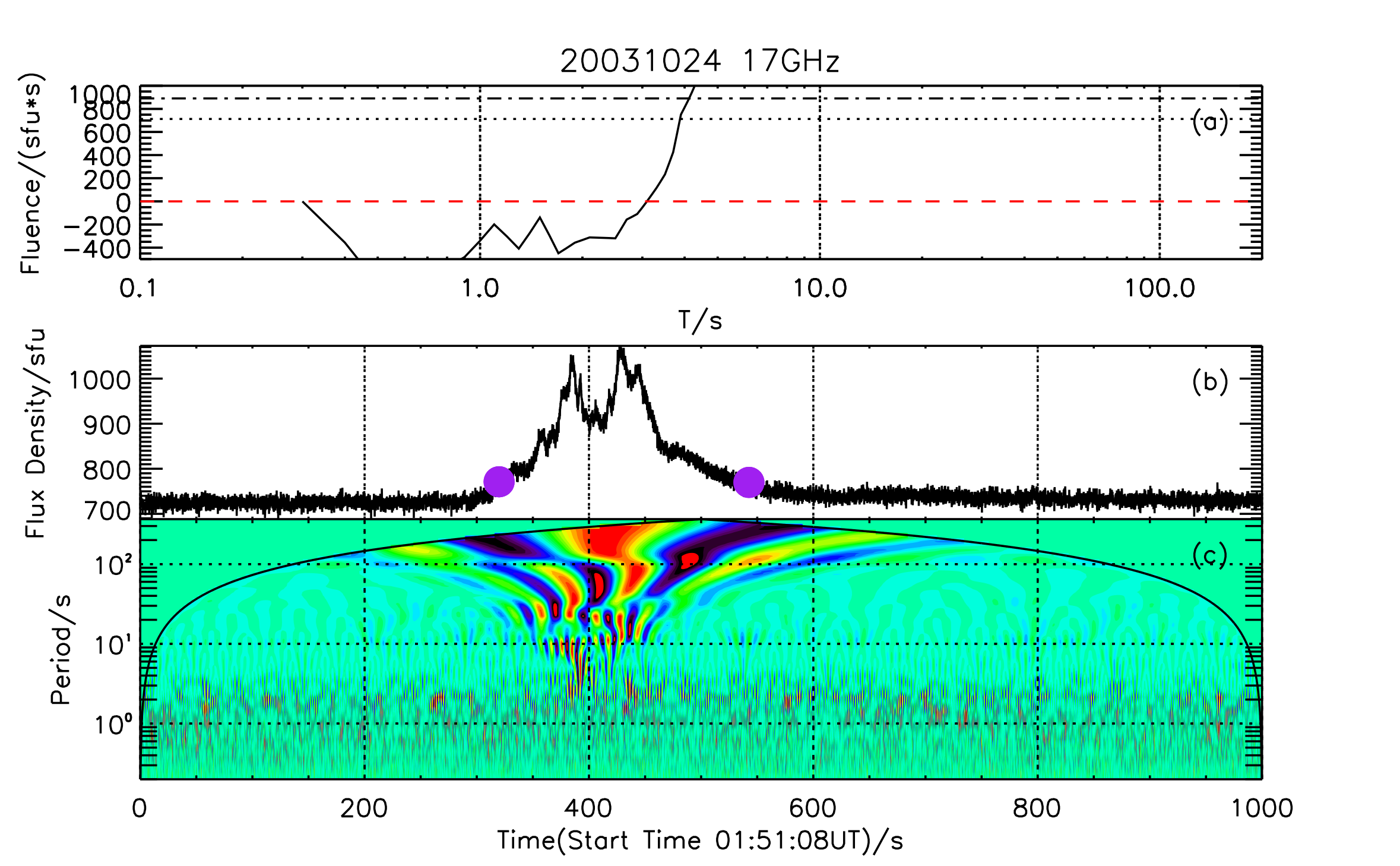}
\end{minipage}  
	\caption{Power spectra (a), light curves (b) and spectrograms of wavelet analyses (c) of a sample of flares. The dotted lines indicate the absolute values of the minima of power spectra. The dotted-dashed lines indicate the selection thresholds.} 
	\label{Fig7}
\end{figure}

Wavelet analysis is a very powerful and standard tool to investigate variabilities in a broad range of timescales(\citealt{2017ApJ...836...84D, 2009A&A...493..259I, 2011A&A...533A..61G, 2015ApJ...798..108I}). Since for most time series, there are more power at longer timescales, the variability at short timescales may not be evident. In order to enhance signals at short timescales, we normalize the wavelet power spectrogram to the maximum amplitude of each frequency(\citealt{2012ApJ...754L..28L}). Figure \ref{Fig6} gives an example and the corresponding power spectra of different intervals obtained with equation \ref{Pspec1}, which shows that our power spectral method can readily capture variations on short timescales. The normalized wavelet spectrogram shows that there are significant flux variations up to the temporal resolution of the observation during two periods of the flare indicated by the red squares. The bottom panels show that the power spectra at short timescales depend on whether these two periods are included in the analysis. The dotted-dashed line indicates a threshold we adopted to identify flares with significant short timescale flux variations. To determine the shortest timescale with significant flux variations, we require that its power be greater than the absolute value of the minimum of the power spectra by 0.5\% of the energy of the gradual component shown in Figure \ref{Fig4} as indicated by the dotted-dashed lines in Figures \ref{Fig6} and \ref{Fig7}. For this event, the shortest timescale with significant variation is about 0.6 second, which is a very conservative value for the slow rise of the power with T. Figure \ref{Fig7} gives a few more examples of such an analysis. For the two flares in the top row, the spectrograms of wavelet analysis show that there are evident short timescale flux variations. Our power spectral method reveals significant flux variation below 0.4 s. For the two flares in the middle row, there is no evident signal in the wavelet spectrograms. We find a flux variation timescale longer than 3 s. The two flares in the bottom row show that there can still be negative power even the power at 0.3 s is set to be zero.

\begin{figure}[htb]
\begin{minipage}{15cm} 
\centering
\includegraphics[width=15cm]{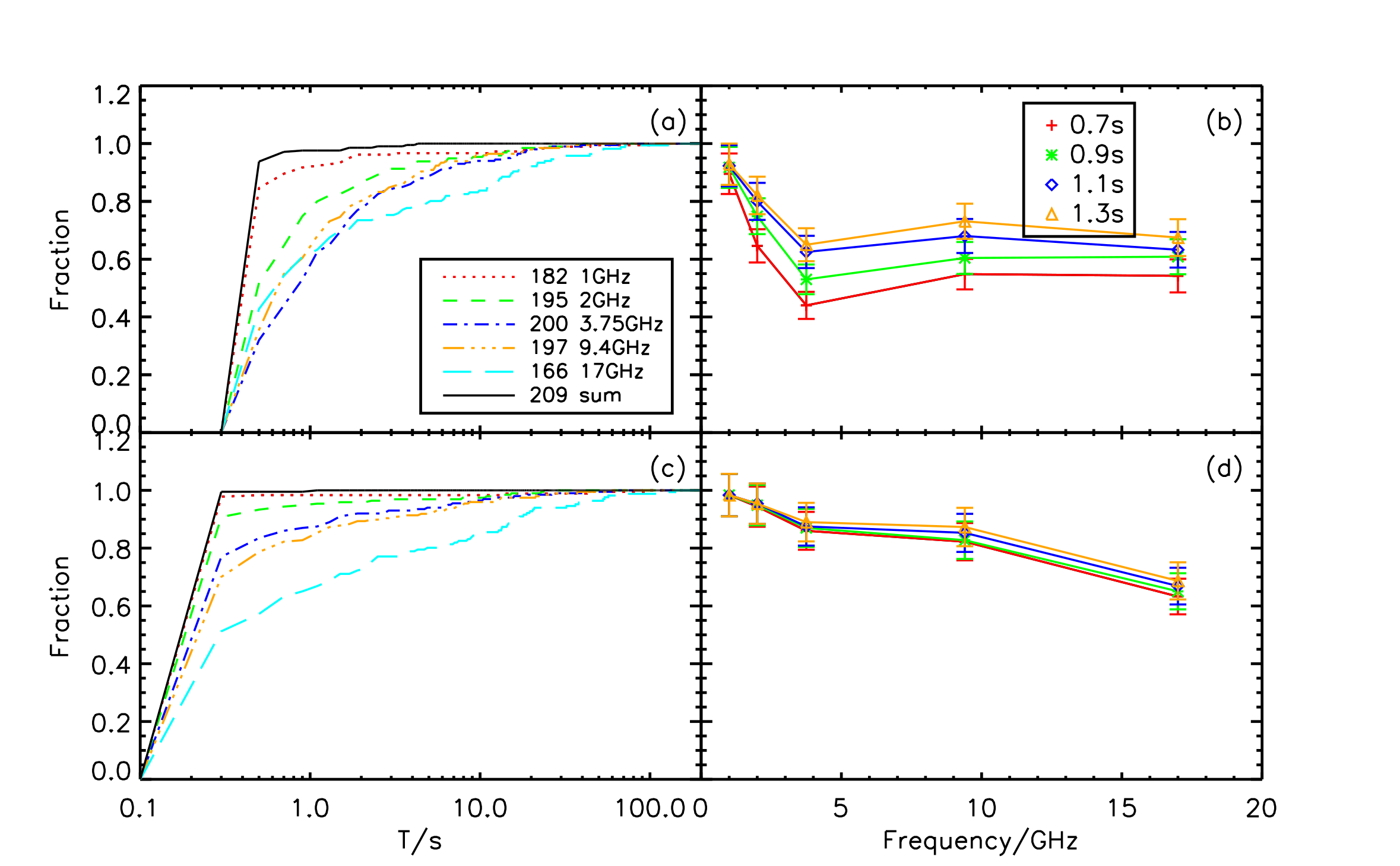}
\end{minipage}
\caption{Dependence of the fraction of flares with their spectral power exceeding the threshold. The upper and lower panels are for two different methods for calculation of the power spectrum. See text for details.} 
\label{Fig9}
\end{figure}

Figure \ref{Fig9} shows the fraction of flares with power at different timescale T exceeding the threshold. The top two panels are obtained with the power spectra calculated with equation \ref{Pspec1}, where the power is set to be zero at T $=$ 0.3 s. The bottom two panels are for those obtained with equation \ref{Pspec0}. The left panels show that more than 90\% of events at 1 GHz have significant short timescale flux variations.This fraction decreases monotonically with the increase of frequency in the low-left panel. The derivatives of these curves show the distribution of shortest timescales with significant flux variations. The lower-left panel shows that essentially all flares have significant flux variation at one of the observation frequencies on the shortest timescale of $0.3$ s. Since the power is set to zero at 0.3 s for the top panels, the fractions are always lower than those in the low panels. In general, this fraction decreases quickly from 1 to 3.75 GHz and then levels off toward very high frequencies (Right panels). 

There appears to be a dip at 3.75 GHz in the top-right panel. In comparison with the low-right panel, it implies that the power spectrum at 3.75 GHz increases more gradually with the increase of timescale than other frequencies below 1 second. This may be caused by a transition from coherent emission dominated impulsive emission at lower frequencies to gyro-cyclotron dominated emission above 3.75 GHz. It may also suggest that there are more spiky bursts near 3.75 GHZ while short timescale signals at other frequencies are dominated by type III kind frequency drift. Statically study of radio spectrogram from 1 to 10 GHz will be able to clarify this issues.  

\section{Correlation with X-rays}
\label{sect:discussion}

\begin{figure}[h]
\begin{minipage}{15cm}
\includegraphics[width=15cm]{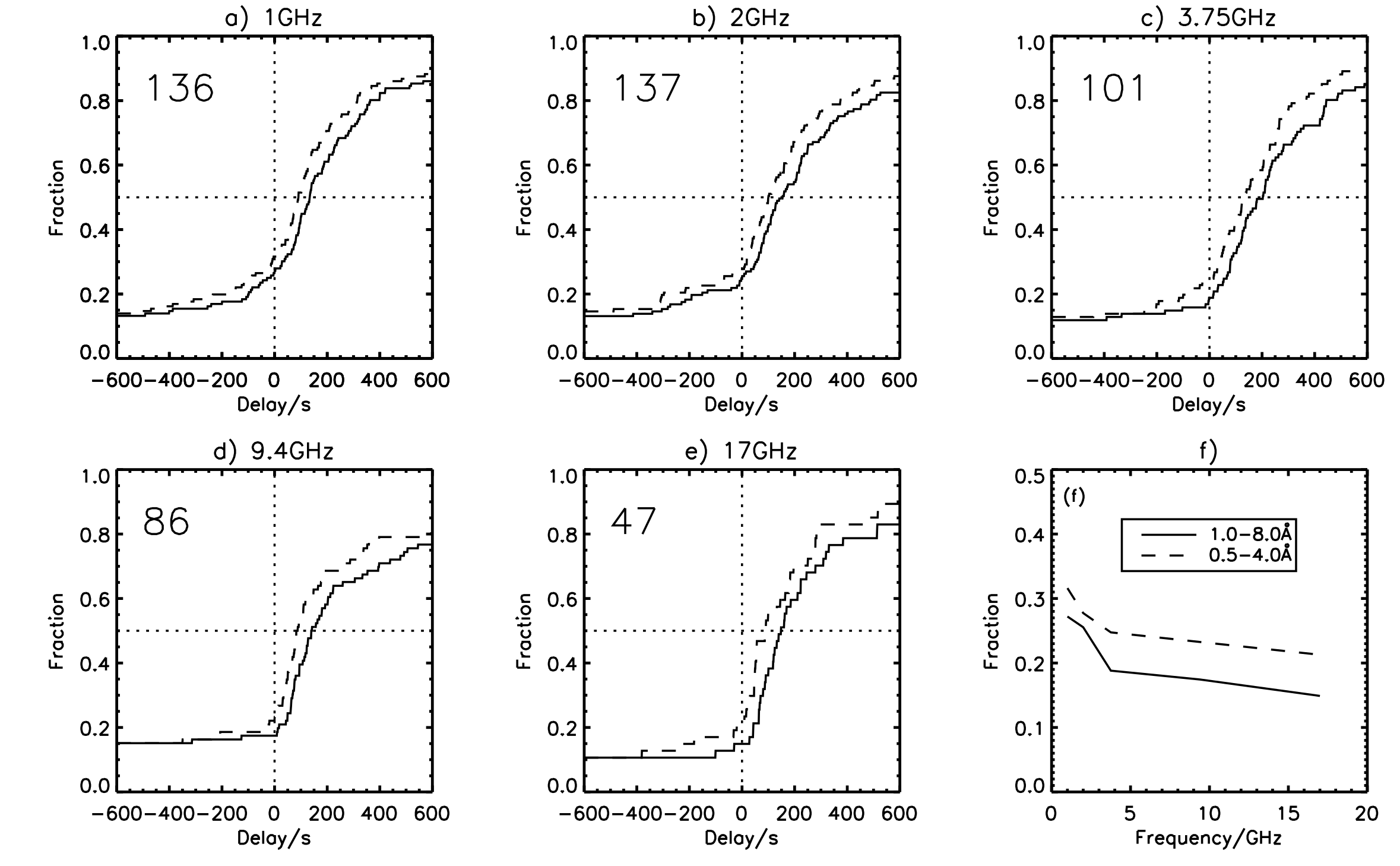}
\end{minipage}
\caption{The distribution of time delay between the peaks of the impulsive component and the soft X-ray fluxes of the GOES satellites. Panel f shows the fraction of events with the GOES fluxes peak ahead of the impulsive radio emission.}
\label{Fig10}
\end{figure}

We also analyse the correlation between radio light curves and X-ray light curves obtained with the GOES satellites. We only consider radio light curves with evident impulsive emission and their peak time can be reliably determined. In total, we find 136, 137, 101, 86, and 47 events for 1, 2, 3.75, 9.4, and 17 GHz, respectively. For these light curves, we obtain their impulsive components as in Section \ref{sect:Method} and study the time delay between the peaks of these impulsive components and the corresponding peaks of the X-ray fluxes observed with the GOES. Figure \ref{Fig10} shows that for most flares, the impulsive component in the radio band peaks ahead of the X-ray emission. Panel f shows that this fraction increases with the increase of frequency. Since the high-energy channel of GOES usually peaks ahead of the low-energy channel, this fraction is also lower for the high-energy channel. These results imply better correlation between higher frequency radio emission and X-ray, which is consistent with the Neupert effect(\citealt{1968ApJ...153L..59N}).

\section{Conclusion}
\label{sect:conclusion}

By analysing radio light curves of 209 flares obtained with the NoRP, we find that emission at 1 GHz is distinct from the other channels. The distribution of the peak flux density of the impulsive component at 1 GHz peaks at a few tens of sfu while at other frequencies this value appears to be below the level of the corresponding background. Distribution of the duration of the impulsive component peaks at a few tens of seconds at 1 GHz and this value shifts gradually to a few seconds with the increase of frequency to 17 GHz. The maximum flux density of the impulsive emission is also the highest at 1 GHz. A value of $10^5$ sfu corresponds to a brightness temperature more than a few tens of billion Kelvin (see Eq.\ref{brightT}). The impulsive 1 GHz emission is therefore dominated by coherent emission processes. Such coherent emission becomes less prominent with the increase of frequency. Above 4 GHz, the impulsive emission appears to have a different origin with brightness temperatures comparable to the corona values. These results are consistent with the scenario that emission at 1GHz is dominated by coherent processes triggered by magnetic reconnection while at high frequencies, the emission is dominated by gyro-synchrotron processes of high-energy electrons(\citealt{2005A&A...440L..59F}).

Considering the resolution of 0.1s of NoRP, one expects even higher brightness temperatures with better temporal resolution (\citealt{2007ApJ...671..964T, 2010Ap&SS.325..251T, 2010ApJ...723...25T}). Such coherent emission is usually attributed to high-energy electrons. In particular, Type III radio bursts have been attributed to beams of nearly relativistic electrons (\citealt{1987SoPh..111..113A, 2008SoPh..253..117T, 2000A&A...360..715K, 2019ApJ...885...90T}). Electron beams can be produced in the acceleration process of magnetic reconnection and both upward and downward beams are expected. The latter can produce the so-called reverse Type III bursts with bulk of the emission drifting toward higher frequencies. Our statistical study of the impulsive radio emission suggests that most reconnection may occur in regions with a plasma frequency higher than 1 GHz, which is consistent with a few cases studied by \citealt{2016ApJ...819...42T}. Future studies of impulsive radio emission at more frequencies from 1 to 10 GHz are warranted and may be compared with detailed modelling of drifting pairs (\citealt{2011ApJ...738L...9L}).

We also find that the energy of the impulsive component is usually more than one order of magnitude lower than the gradual component. The fact that the energy of the impulsive component increases slowly with the increase of the product of its peak flux density and duration suggests that its flux density distribution decreases sharply with the increase of the flux density, the power-law scaling of these two quantities imply similarity of flares of different magnitude. The energy of the gradual component increases linearly with the product of its peak flux density and duration, which is distinct from the impulsive emission.

\normalem
\begin{acknowledgements}
This work is partially supported by the International Partnership Program of Chinese Academy of Sciences (No. 114332KYSB20170008) and the International Cooperation and Exchange Project of National Natural Science Foundation of China (No. 11761131007).
\end{acknowledgements}
  
\bibliographystyle{raa}
\bibliography{ref}

\section{Appendix A}
\label{sec:App A} 
\begin{figure}[hbt]
\begin{minipage}{15cm}
	\includegraphics[width=15cm]{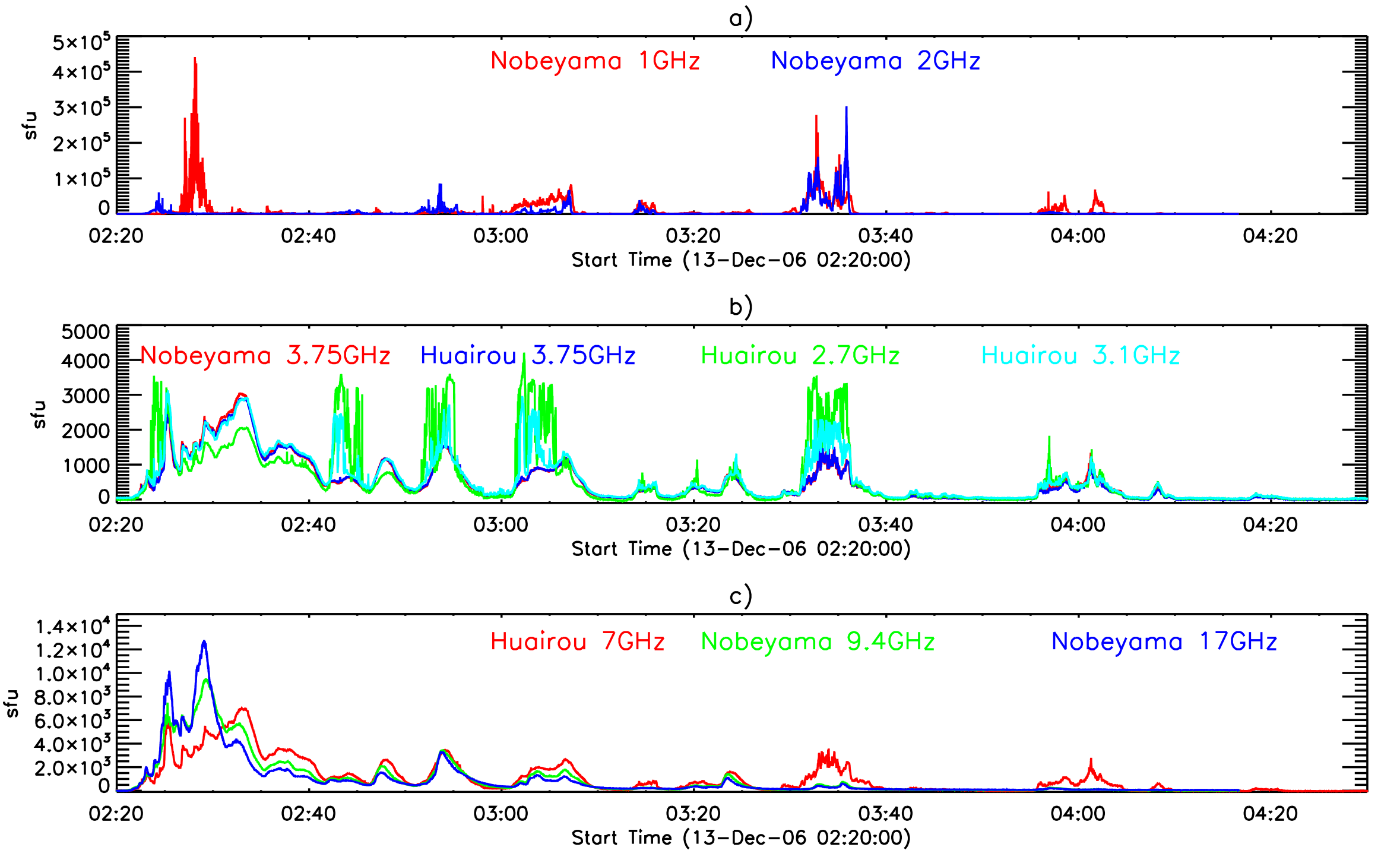}
\end{minipage}
	\caption{Light curves of a flare occurred on Dec 13 2006 at several frequencies. }
	\label{Fig11: Appendix A1}
\end{figure}

\begin{figure}
\begin{minipage}{15cm}
	\includegraphics[width=15cm]{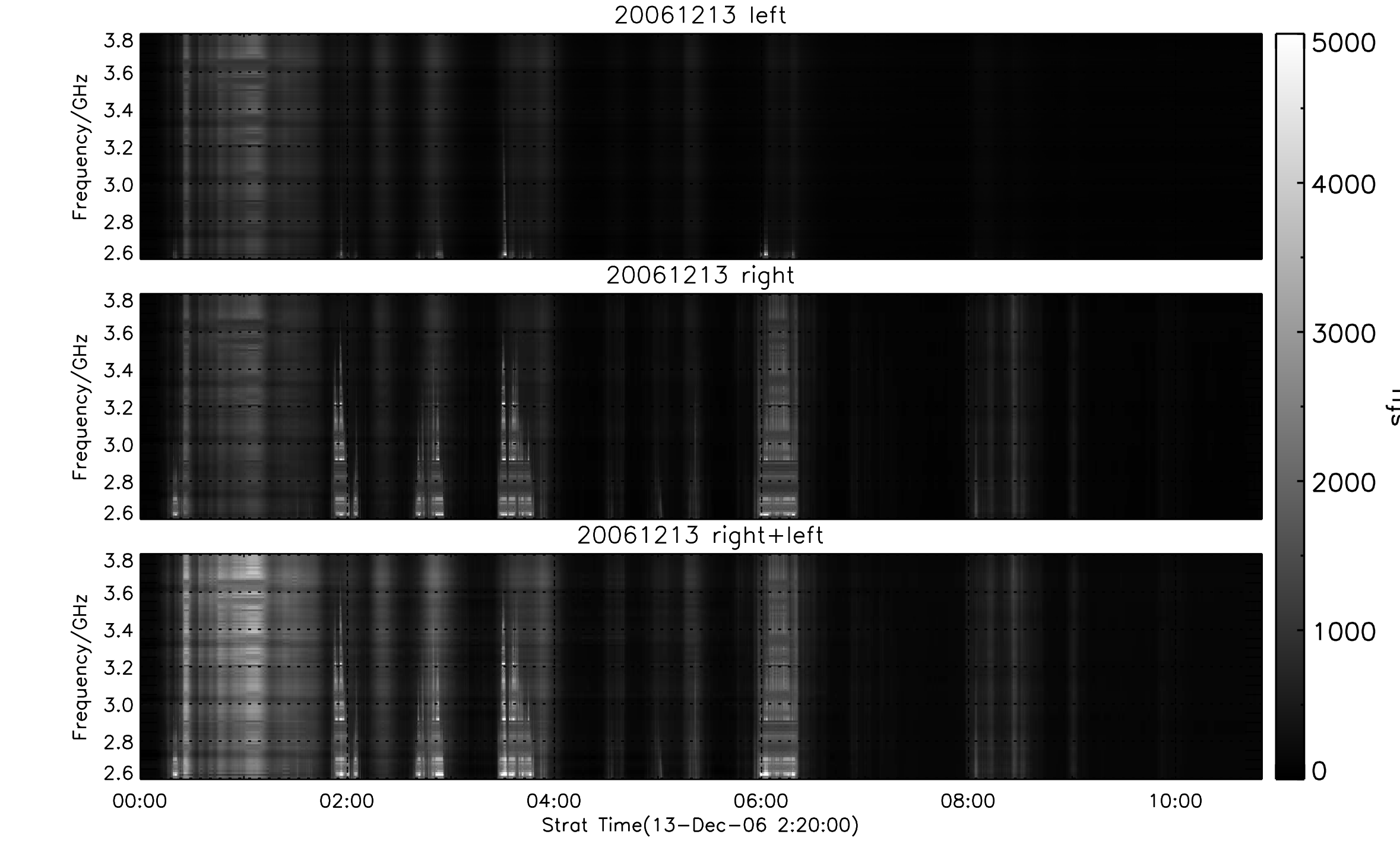}
\end{minipage}
	\caption{Radio spectrogram of the flare on Dec 13 2006 obtained with the Huairou solar radio telescope.}
	\label{Fig12: Appendix A2}
\end{figure}

\begin{figure}
\begin{minipage}{15cm}
	\includegraphics[width=15cm]{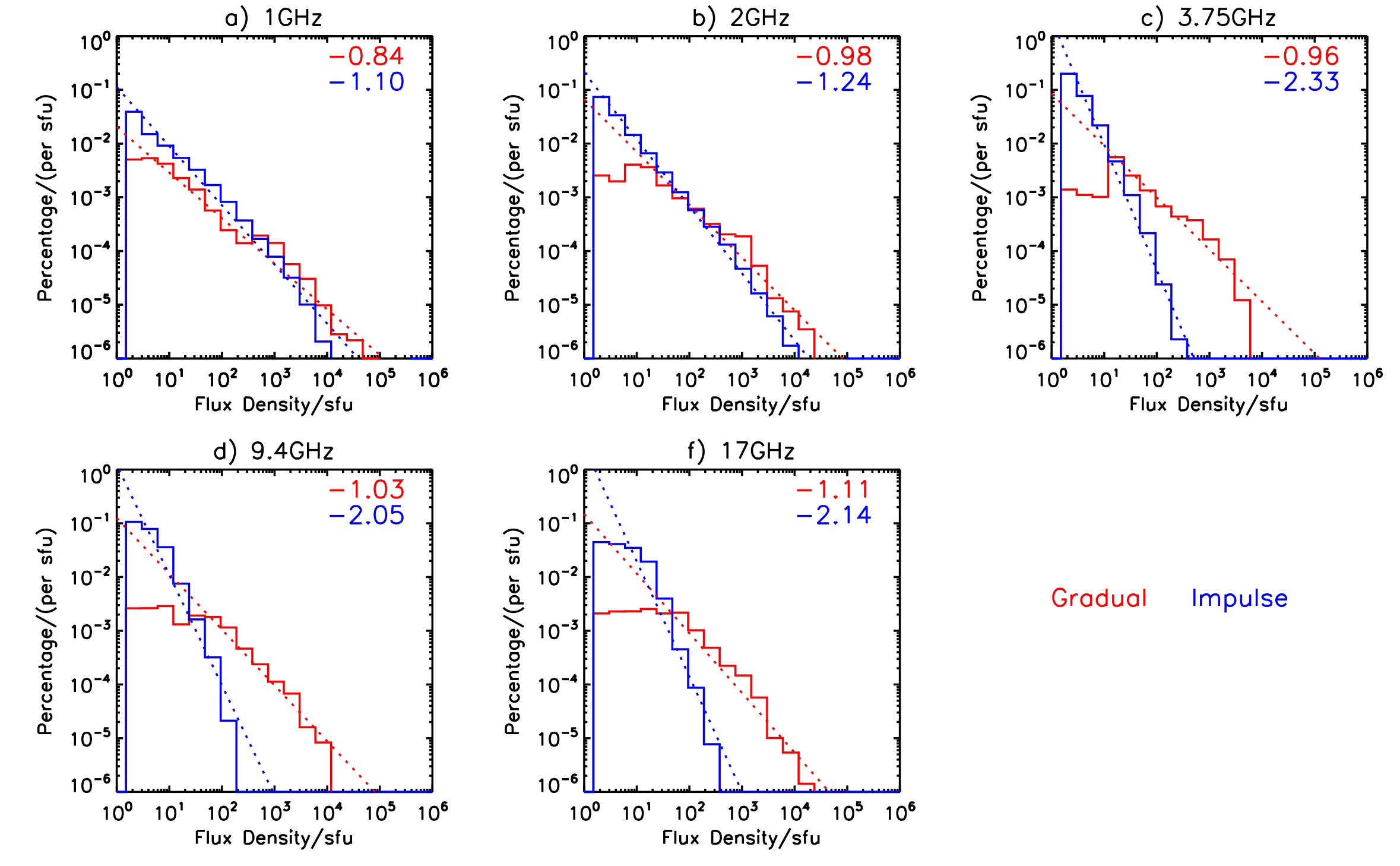}
\end{minipage}
	\caption{The flux density distribution of the impulsive and gradual components.}
	\label{Fig13: Appendix A3}
\end{figure}

Here we show a flare with very high impulsive flux densities, which are indicated by squares in Figures \ref{Fig2}, \ref{Fig3}, \ref{Fig4} and red signs in Figure \ref{Fig5}. Figure \ref{Fig11: Appendix A1} shows that the 1 and 2 GHz emissions are dominated by the impulsive component. The emission is more gradual at higher frequencies. Figure \ref{Fig12: Appendix A2} shows the spectrogram of the polarized and total emission. Figure \ref{Fig13: Appendix A3} shows that impulsive component has a steeper distribution than the gradual component, which agrees to results in Figure \ref{Fig5}.

\end{document}